# Distinguishing impurity-induced bound states from Majorana-like zero-energy peaks in strained $CsCa_2Fe_4As_4F_2$ by scanning tunneling microscopy

Mingzhe Li[1], Jiashuo Gong[1], Huaxun Li[2], Jiakang Zhang[1], Yuanji Li[1], Ruotong Yin[1], Shiyuan Wang[1], Guanghan Cao[2], Dong-Lai Feng[3,*], Ya-Jun Yan[1,3,*]

[1]*Hefei National Research Center for Physical Sciences at the Microscale and Department of Physics, University of Science and Technology of China, Hefei, 230026, China*
[2]*School of Physics, Zhejiang University, Hangzhou, 310058, China*
[3]*New Cornerstone Science Laboratory, Hefei National Laboratory, Hefei, 230088, China*

**Corresponding authors:** dlfeng@hfnl.cn; yanyj87@ustc.edu.cn

**Iron-based superconductors offer a versatile platform for exploring topological superconductivity and Majorana zero modes (MZMs), with experimental confirmations in Fe(Te,Se), (Li,Fe)OHFeSe and $CaKFe_4As_4$ at ambient pressure, as well as in LiFeAs under local strain. The related properties in other iron-based superconductors still need to be explored, especially under the application of local strain. In this study, we conduct scanning tunneling microscopy/spectroscopy measurements on $CsCa_2Fe_4As_4F_2$ crystals under unidirectional local strain. A fully developed superconducting gap with multiple pairs of coherence peaks are observed, and the gap sizes can be significantly modulated by local strain. Spectroscopic measurements on various types of defects including the nonmagnetic Cs-site vacancies consistently reveal pair-breaking effects. These phenomena support a fully gapped multiband superconductivity scenario with sign-changing. Notably, a sharp zero-energy conductance peak (ZECP) is universally observed on a particular type of defects by using a metallic tip, resembling the MZMs observed at interstitial Fe atoms in Fe(Te,Se) [Nat. Phys. 11, 543 (2015)]. However, by using a superconducting tip to enhance energy resolution as well as by studying the ZECP evolution as functions of magnetic field and tunneling transmissivity, we demonstrate that the ZECP originates from nearly degenerate Yu-Shiba-Rusinov states rather than MZMs. Our study not only provides more insights into the superconducting pairing symmetry of $CsCa_2Fe_4As_4F_2$, but also establishes systematic experimental methods for identifying weak impurity state signals and discerning the physical origins of ZECPs.**

## Introduction

Beyond high-temperature superconductivity, iron-based superconductors have emerged in recent years as a promising platform for exploring topological superconductivity and Majorana zero modes (MZMs). For instance, MZMs were directly observed at the magnetic vortex core, interstitial Fe atoms, line defects or domain walls in Fe(Te,Se), (Li,Fe)OHFeSe, $CaKFe_4As_4$ and monolayer $FeSe/SrTiO_3$ at ambient pressure [1-10]; while in LiFeAs, topological superconductivity can be realized by applying local strain to adjust the topological surface states to $E_F$ [11-13]. Moreover, recent dynamic mean field theory (DMFT) calculations reveal that when strong correlation effects are taken into account, more iron-based superconductors will have topological surface states at $E_F$, thereby giving rise to topological superconductivity. The suggested systems include the 1111 and

122 families [14]. However, the exposed surfaces of 1111 and 122 families after cleavage are polar surfaces with strong trivial surface states, and the superconducting gap is suppressed or almost invisible on them [15-17], which greatly hinders the measurements of intrinsic electronic properties, and the evidence for topological superconductivity and MZMs is still lacking.

$ACa_2Fe_4As_4F_2$ (A = K, Rb, Cs) is formed by the alternating intercalation of CaFeAsF and $AFe_2As_2$, which possesses a relatively high superconducting transition temperature ($T_c$ ~ 28-33 K). Previous scanning tunneling microscopy/spectroscopy (STM/STS) studies have revealed a cleavage plane of nonpolar Cs layer and a good superconducting gap on it, offering a new candidate platform to study the interplay between high-temperature superconductivity and band topology. In terms of topological properties, there is no report yet on topological surface states and topological superconductivity without strain, but the situation under local strain is still unclear. Furthermore, we notice that there are still some controversies in previous experiments regarding the superconducting pairing symmetry of $ACa_2Fe_4As_4F_2$. Early muon-spin rotation (μSR), specific heat, and point-contact spectroscopy experiments supported nodal superconducting gap structures in all three $ACa_2Fe_4As_4F_2$ (A = K, Rb, Cs) compounds [18-23]. However, ultralow-temperature thermal conductivity, high-resolution laser-based angular resolved photoemission spectroscopy (ARPES), and STM/STS experiments revealed a fully gapped multiband scenario in $ACa_2Fe_4As_4F_2$ (A = K, Cs) [24-27]. To further understand the superconducting pairing symmetry of $ACa_2Fe_4As_4F_2$ and to explore possible topological superconductivity, in this study, we conduct STM/STS measurements on $CsCa_2Fe_4As_4F_2$ crystals under unidirectional local strain. A fully-developed superconducting gap with multiple pairs of coherence peaks are observed, and the gap sizes can be significantly modulated by local strain. Spectroscopic measurements on various types of defects including the nonmagnetic Cs-site vacancies consistently reveal pair-breaking effects and in-gap bound states. These phenomena support a fully gapped multiband superconductivity scenario with sign-changing. Notably, a sharp zero-energy conductance peak (ZECP) is universally observed on a particular type of defects, resembling MZMs observed at interstitial Fe atoms in Fe(Te,Se) system [1]. However, through systematic studies from several aspects, we demonstrate that the ZECP originates from impurity-induced nearly degenerate in-gap bound states rather than MZMs.

## Methods

$CsCa_2Fe_4As_4F_2$ single crystals used in this study were grown by the self-flux method as described in Ref. [28]. For STM study, the $CsCa_2Fe_4As_4F_2$ crystals were mechanically cleaved at ~80 K in ultrahigh vacuum with a base pressure better than $1 \times 10^{-10}$ mbar and then immediately transmitted into the STM module. STM/STS experiments were carried out in a UNISOKU cryogenic STM at $T$ = 80 mK. PtIr tips were used after being treated on clean Au (111) substrates. To make a superconducting tip, the metallic PtIr tips were indented into a Pb (111) surface cleaned by repeated sputtering and annealing [29]. The Pb-coated superconducting tip was calibrated on Pb (111) surface to obtain the superconducting gap information before measuring $CsCa_2Fe_4As_4F_2$ samples, as discussed in Appendix A. The d$I$/d$V$ spectra were collected using a standard lock-in technique with a modulation frequency of 973 Hz and a modulation amplitude of $\Delta V$ = 0.02 ~ 1 mV.

## Experimental Results
### A. Determination of various types of defects

Figure 1(a) shows the crystal structure of $CsCa_2Fe_4As_4F_2$, which can be easily cleaved along the Cs atomic layers indicated by the magenta plane, leaving behind a reconstructed Cs-terminated surface as previously reported [27]. Figures 1(b) and 1(c) present representative topographic images of the exposed surface over different field of view (FOV), and the inset of Fig. 1(b) resolves clearly the surface Cs atomic lattice with a $\sqrt{2} \times \sqrt{2}$ reconstruction. Based on this information, the orientation of the Fe–Fe lattice can be determined, which aligns with the reconstructed Cs atomic lattice and is marked out in the lower-left corner of each panel. Notably, the broad-view topographic image in Fig. 1(b) reveals obvious unidirectional wrinkles across the surface, as highlighted by the height line profiles (lower panel of Fig. 1(b)), which deviate from one of the Fe-Fe directions, here referring to the Fe1 direction, by approximately 18°~20°. These wrinkles indicate the presence of unidirectional local strain which likely arises from the lattice mismatch between CaFeAsF and CsFe$_2$As$_2$ layers as well as lattice relaxation during crystal cleavage.

Figure 1(d) displays representative superconducting gap spectra collected above and below the wrinkles, as marked out by the black and red dots in Fig. 1(b). Both spectra exhibit four pairs of characteristic coherence peaks between 3-10 meV and nearly zero density of states (DOS) around $E_F$, revealing the fully-gapped multiband superconductivity in CsCa$_2$Fe$_4$As$_4$F$_2$. This is consistent with previous ARPES results that revealed nodeless superconducting gaps on the five holelike Fermi surface sheets around the Brillouin zone center as well as a tiny pocket surrounded by four strong spots around the zone corner [25,26]. The observed superconducting gaps here originate from different energy bands with distinct orbital characteristics. It is worth noting that a fully developed superconducting gap with only one or two pairs of coherence peaks of ~ 5-8 meV was commonly observed in STM studies on strain-free ACa$_2$Fe$_4$As$_4$F$_2$ sample [27,30,31]; while here, the presence of local strain enables more superconducting gaps to be clearly distinguished. This phenomenon should be attributed to the tunneling selectivity of STM measurements on different orbitals. Typically, orbitals with a finite $z$-component (such as d$_{xz}$ and d$_{yz}$) overlap more favorably with the wave function of the STM tip, rendering them easier to be detected. For strain-free regions, STM predominantly probes the d$_{xz}$ and d$_{yz}$ orbital components, while being relatively insensitive to in-plane orbitals like d$_{xy}$. Under local strain such as the wrinkles observed in this study, however, the topography exhibits fluctuations in the $z$ direction, causing those originally in-plane orbitals to develop components along the $z$-axis, which facilitates the detection of their superconducting gaps. Moreover, the strain gradient can further modulate the superconducting gap size, as clearly discerned in Fig. 1(d): compared to the spectrum above the wrinkle (black curve), the sizes of the superconducting gap $\Delta_1$ and $\Delta_2$ for the spectrum (red curve) below the wrinkle decrease, while $\Delta_3$ increases significantly, and $\Delta_4$ remains almost unchanged. The different responses of superconducting gaps to unidirectional strain should be related to the different regulatory effects on energy bands or orbitals.

Besides, there are numerous randomly distributed dark holes and several dark strips (white arrows) in Fig. 1(b). The former corresponds to defect I circled by the black dotted line in Fig. 1(c), which is probably a Cs-site vacancy; while the latter are basically along the Fe1 direction, which are probably contiguous lines of missing Cs atoms. From the higher-resolution topographic image in Fig. 1(c), we can distinguish another two types of defects, namely defects II and III circled by the purple and blue dotted lines, which are characterized by a dumbbell structure consisting of two bright lobes and an elongated bright spot, respectively. To further distinguish these different types of defects and to reveal other unobservable defects in topographic image, we measure d$I$/d$V$ maps within the superconducting gap energy range. The results are shown in Figs. 1(e)-1(h) and Appendix

B, and we identify three typical point defects with distinct in-gap bound states.

1) The first type is defect II as visible in topographic image (Fig. 1(c)). It induces in-gap bound states, which are observable within the energy range of ±2 meV and are the strongest at E = -2 meV. The spatial distribution of the in-gap bound states for all defects of type II are the same, maintaining the $C_2$ symmetry with two bright lobes, but their orientations change with energy (Figs. 1(e)-1(g)).

2) The in-gap bound states induced by defect III are the strongest at E = 2 meV (blue dashed circle), manifested as a bright spot elongated along approximately one As-As direction, which is similar to the pattern in topographic image. However, at E = -2 meV, they exhibit a two-lobed structure with weaker intensity, and the symmetry axis is along the elongation direction of the bright spot.

3) Notably, there are other defects of type IV, which are nearly absent in topographic image but become unambiguous in d$I$/d$V$ map at E = 0 meV (red dashed circles in Figs. 1(c) and 1(f)). Their in-gap states are all $C_2$ symmetric and exhibit dumbbell-shaped spatial distribution. The dumbbells have only one preferred orientation along the Fe1 direction in the whole FOV.

In the absence of strain, both the crystal structure and electronic structure of $CsCa_2Fe_4As_4F_2$ exhibit in-plane fourfold symmetry. Under such conditions, defects at Cs- and As-sites show $C_4$-symmetric features in topographic images and d$I$/d$V$ maps, while the Fe-site defects adopt a $C_2$-symmetric dumbbell-shaped structure with two degenerate orientations along the As–As directions. Such symmetry differences have been widely used to distinguish between Fe-site and As-site defects in iron-based superconductors [32-35]. In our case, the Cs-site defects reside on the topmost cleaved surface and are readily identifiable in topographic images; in contrast, defects II-IV are subsurface features with weak topographic contrast, complicating their identification. Moreover, the spatial distribution of the in-gap bound states of defects II-IV all exhibit strong $C_2$ symmetry; and for each defect type, only one preferred orientation is observed, with no evidence of orientational degeneracy. This is likely closely related to the presence of unidirectional strain, which induces pronounced electronic nematicity along the Fe-Fe directions, as demonstrated by the unidirectional quasiparticle interference patterns in Figs. 1(e)-1(h) and Fig. 7 in Appendix B. Consequently, due to this strain-induced electronic nematicity, it becomes even more difficult to further determine the atomic sites of defects II-IV based solely on the symmetry of spatial distributions of their in-gap bound states. They likely correspond to Fe-site defects, As-site defects, or interstitial Fe atoms located within or between the FeAs layers, and are most probably magnetic in nature [36,37]. Clarifying this issue would require further theoretical modeling and will be not discussed in detail here.

### B. Impurity effects in the superconducting state

Figure 2 displays detailed evolution of the superconducting gap spectrum across different types of defects mentioned above. Although local strain modulates the superconducting gaps, the superconducting gap spectra remain essentially uniform on the scale of a few nanometers in defect-free areas (Fig. 2(a)). This homogeneity provides a reliable foundation for evaluating impurity effects. Figure 2(b) presents the d$I$/d$V$ spectra across an individual Cs-site defect shown in the inset, minimal effect can be observed and the spectra taken atop the defect (red curves) remain almost unchanged relative to those in the defect-free area. Such weak perturbation is consistent with the d$I$/d$V$ maps in Fig. 1, where the Cs-site defects induced variation in local DOS is subtle. This is expected, as Cs atoms are located outside the primary superconducting FeAs planes, thus their vacancies exert weak scattering potential for Cooper pairs. For the line defects of Cs-site vacancies shown in Fig. 2(c), although no obvious in-gap bound states are resolved, the DOS near $E_F$ is lifted significantly on the defect. This is evident from the two overlaid spectra on and off the line defect (marked by blue dots), indicating a local suppression of superconductivity. In stark contrast, defects II-IV show much stronger pair-breaking effects, inducing obvious in-gap bound states as indicated by the blue arrows in Figs. 2(d)-2(f). The strongest in-gap bound states locate at E ≈ -2 meV, 2 meV, and 0 meV for defects II-IV, respectively, consistent with the results shown in Figs. 1(e)-1(g). These phenomena reflect the close association of defects II-IV with FeAs layer, which probably locate in the FeAs layer thus can induce stronger scattering of Cooper pairs compared to the Cs-site defects.

To further determine the impurity effect of Cs-site defects, we use a superconducting Pb tip for spectroscopy measurements, which can significantly enhance the energy resolution, thereby enabling the discrimination of delicate spectroscopic signals [29,38-40]. The typical superconducting gap spectra acquired in defect-free area and atop a Cs-site defect are shown in Figs. 3(a) and 3(b), both exhibiting sharp symmetric coherence peaks at $\pm\Delta_{tip}$, $\pm(\Delta_{tip}+\Delta_1)$, $\pm(\Delta_{tip}+\Delta_2)$ and $\pm(\Delta_{tip}+\Delta_3)$ at approximately ±1.3 meV, ±5.8 meV, ±7.1 meV and ±8.9 meV, respectively. However, the spectra differ significantly within the energy range between $\Delta_{tip}$ and $\Delta_{tip}+\Delta_1$ — it is relatively clean in Fig. 3(a), while obvious symmetric in-gap bound states emerge in Fig. 3(b) (blue arrows). To further clarify this point, these spectra are deconvoluted to extract the intrinsic spectra for $CsCa_2Fe_4As_4F_2$. The deconvolution process is described in Appendix C and the obtained results are shown in Figs. 3(c) and 3(d). Consistent phenomena are revealed: the spectrum in the defect-free area exhibits a clean U-shaped superconducting gap, the slight lift of the in-gap DOS can be attributed to minor contribution of Andreev reflection effect due to the large tunneling current used to acquire the high-resolution data (Fig. 3(c)); while the spectrum atop Cs-site defect features obvious in-gap bound states (blue arrows in Fig. 3(d)). These phenomena suggest that the Cs-site defects do have an impact on the superconducting state of $CsCa_2Fe_4As_4F_2$, consistent with the results on line defect of Cs-site vacancies (Fig. 2(c)).

These impurity effects provide critical new insights into the nature of superconducting pairing symmetry in $CsCa_2Fe_4As_4F_2$. All the defects are destructive to the superconducting state, especially the Cs-site defects that are nonmagnetic scatterers, strongly supporting a sign-changing pairing symmetry, which aligns with the generally accepted $s\pm$ pairing symmetry in iron-pnictide superconductors [27,41-44].

### C. Response of the ZECP on defect IV to magnetic field and tunneling transmissivity

Another intriguing finding of this study is the presence of a sharp ZECP on defect IV (Fig. 2(f)). To confirm the universality, we measure the zero-energy d$I$/d$V$ map for several sample regions, and one of them is displayed in Fig. 4(a). Almost all the defects of type IV manifest similarly as bright dumbbells oriented along the Fe1 direction. Figure 4(b) shows the d$I$/d$V$ spectra collected on eleven defects of type IV, as numbered in Fig. 4(a); All of them consistently exhibit a sharp ZECP, confirming that this is a robust characteristic of defect IV. As shown in Fig. 4(c), the intensity of the ZECP can be even greater than that of the coherent peaks, and its full width at half maximum (FWHM) can be narrowed down to 0.22 meV. Such robust ZECP reminds of MZMs observed at interstitial Fe atoms in (FeTe,Se) superconductors, but possesses an even narrower FWHM than previous reports of ~ 0.3-0.6 meV [1,2,4,11]. If it is truly topological MZMs, it must be strictly at zero energy and exhibit a strong resistance to magnetic field and tunneling transmissivity [1,5,45,46]. We have conducted more experiments to verify this.

Firstly, we utilize the high energy resolution of the superconducting tip to distinguish whether this ZECP is strictly at zero energy. If it is, the coherent peaks of the superconducting tip located at $\pm\Delta_{tip}$ should maintain particle-hole symmetry in both energy and intensity [40,47]; otherwise, they do not need. The typical spectrum measured on defect IV using a superconducting Pb tip is shown in Fig. 4(d). It is obvious that the intensities of the pair of coherent peaks at $\pm\Delta_{tip}$ differ significantly, and there is also another pair of sharp side peaks with symmetric energy but asymmetric intensity (blue arrows), which was not observed by using a normal metallic tip (Fig. 4(c)). Figure 4(e) shows the spectrum after deconvolution, its overall shape is similar to that obtained with a metallic tip (Fig. 4(c)), but more details are present around $E_F$, as highlighted in the inset of Fig. 4(e). Firstly, the contour of the so-called ZECP is not symmetric relative to $E_F$, and the energy with the strongest peak intensity deviates from zero energy (red arrow); Secondly, a pair of new side peaks emerges around $E_F$ (blue arrows). This behavior is inconsistent with MZMs, and instead points to two pairs of in-gap bound states, arising from multi-orbital physics/characteristics. One pair of the in-gap bound states has an energy infinitely close to zero, being nearly degenerate, and the other pair is also quite close to zero energy. When using a metallic tip, due to insufficient energy resolution, they merge together, resulting in a broad peak located at zero energy.

Secondly, we study the response of the so-called ZECPs to vertical magnetic fields. Although magnetic vortices appeared under magnetic fields, all impurity states measured here were deliberately chosen from regions far away from vortices to avoid any possible interaction between vortices and impurity states. As shown in Fig. 4(f), most of the ZECPs are only slightly suppressed at 2 T, but at 5 T, their intensities are strongly suppressed and their FWHMs broaden significantly. Figures 4(g) and 4(h) display statistical analysis of FWHM and peak intensity for multiple defects of type IV under different conditions, further confirming gradual suppression of peak intensity and broadening of FWHM with increasing magnetic field. This behavior also contradicts MZMs that are inert to magnetic field.

Thirdly, we investigate the evolution of the so-called ZECP as a function of tunneling transmissivity. The transmissivity of tunneling barrier, defined as $G_N = I_t/V_b$, can be changed by controlling the tip-sample distance. We gradually increase the tunneling current $I_t$ at a fixed $V_b$ to reduce the tip-sample distance, and to increase $G_N$. During this process, the electrostatic force from the approaching tip will affect the coupling between the defects and the superconductor [7,48]. In this condition, MZMs will not move or split, and the zero-energy conductance will gradually approach the quantized conductance value of $G_0 = 2e^2/h$ [5,46,49]. Our results on defect IV are displayed in Fig. 5, revealing a behavior that is fundamentally incompatible with the prediction of MZMs. With increasing $G_N$, the ZECP gradually splits into two energy-symmetric branches that move away from $E_F$ (Figs. 5(a) and 5(b)), which can be better visualized in the corresponding color maps (Figs. 5(c) and 5(d)). Such tunneling transmissivity-dependent evolution resembles the characteristic behavior of impurity-induced in-gap bound states, whose energies are initially nearly degenerate and shift away from each other with increasing coupling [48,50,51].

### Discussion and conclusion

In this study, we systematically investigate the superconducting properties of $CsCa_2Fe_4As_4F_2$. In regions subjected to unidirectional local strain, a fully developed superconducting gap with multiple coherence peaks is observed, consistent with the results from ARPES measurements. Unidirectional strain simultaneously modulates the superconducting state, but affects differently on the superconducting gaps in different bands with different orbital characters. Moreover, it breaks the fourfold rotational symmetry of electronic states, inducing significant electronic nematic characteristics, which are clearly reflected in the quasiparticle scattering patterns and the spatial distribution of impurity states. Furthermore, this study systematically examines the influence of different types of defects on superconductivity: when using a metallic tip, all defects except the Cs-site defect (defect I) exhibit pronounced pair-breaking effects; after employing a superconducting tip to enhance energy resolution, in-gap bound states induced by Cs-site defect are successfully detected. These phenomena collectively point to a sign-changing superconducting pairing form, which aligns with the widely accepted $s\pm$ pairing scenario in iron-pnictide superconductors. Notably, a sharp ZECP is commonly observed on defect IV, which is confirmed to originate from impurity-induced in-gap bound states rather than MZMs, through detailed characterization from multiple aspects. This work not only reveals the effective regulatory role of local strain on superconductivity and electronic states, but also establishes systematic experimental methods for identifying weak impurity state signals and discerning the physical origins of ZECPs. It provides a new research perspective for exploring the pairing symmetry and topological properties of unconventional superconductors under local strain.

## Acknowledgments

This work is supported by the National Natural Science Foundation of China (Grants No. 12374140, No. 12494593, No. 12074363), the National Key R&D Program of the MOST of China (Grants No. 2023YFA1406304), the Innovation Program for Quantum Science and Technology (Grant No. 2021ZD0302803), the New Cornerstone Science Foundation.

## Data availability

The data supporting the findings of this study are available within the article. All the raw data generated in this study are available from the corresponding author upon request.

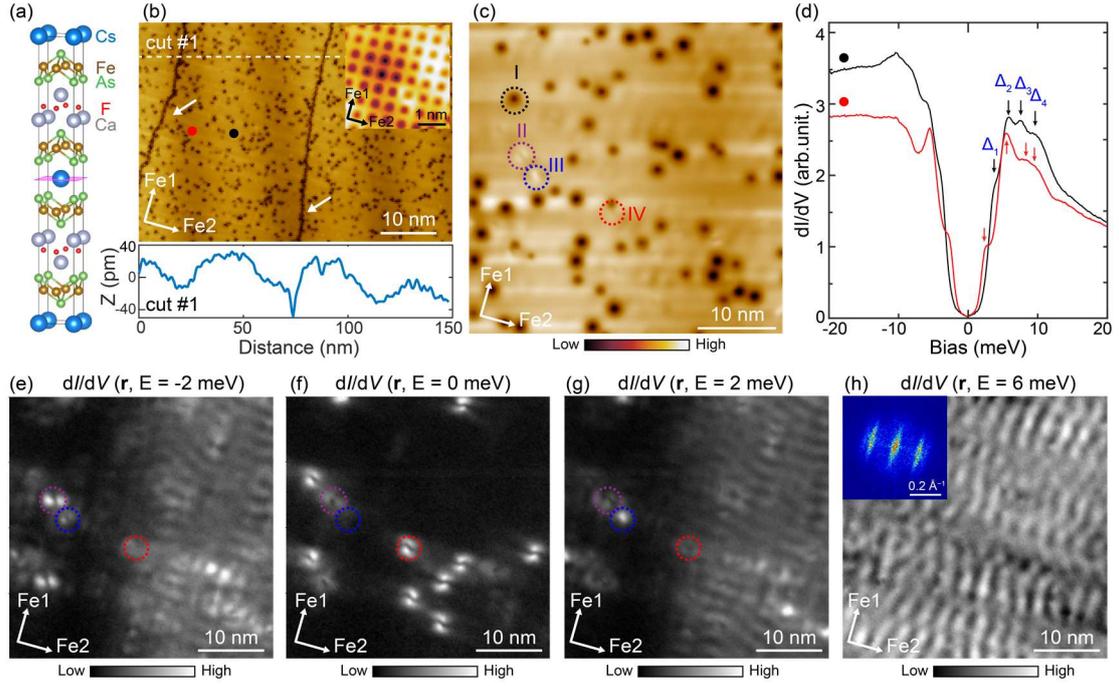

**FIG. 1. Determination of various types of defects in CsCa$_2$Fe$_4$As$_4$F$_2$ with unidirectional local strain.** (a) Crystal structure of CsCa$_2$Fe$_4$As$_4$F$_2$. (b) Typical topographic image after cleavage. The inset shows the atomically resolved topographic image with a $\sqrt{2} \times \sqrt{2}$ surface reconstruction, and the lower panel displays height line profiles taken along cut #1. The white arrows show the line defects of Cs vacancies. (c) Topographic image within a small FOV, highlighting four types of point defects (labeled I-IV). (d) Representative superconducting gap spectra collected on/off wrinkles as marked by the black and red dots in panel (b), revealing multiple coherence peaks as highlighted by the black and red arrows. (e)-(h) d$I$/d$V$ maps under E = -2 meV, 0 meV, 2 meV, and 6 meV, respectively, collected in the same FOV of panel (c), illustrating energy-dependent electronic states distribution and defect states. The FFT image of panel (h) is shown in the inset. Measurement conditions: (b) $V_b$ = 500 mV, $I_t$ = 20 pA; inset of (b): $V_b$ = -1.2 V, $I_t$ = 8.5 nA; (c) $V_b$ = 20 mV, $I_t$ = 100 pA; (d) $V_b$ = 20 mV, $I_t$ = 300 pA, $\Delta V$ = 0.2 mV; (e)-(h) $V_b$ = 20 mV, $I_t$ = 300 pA, $\Delta V$ = 0.3 mV.

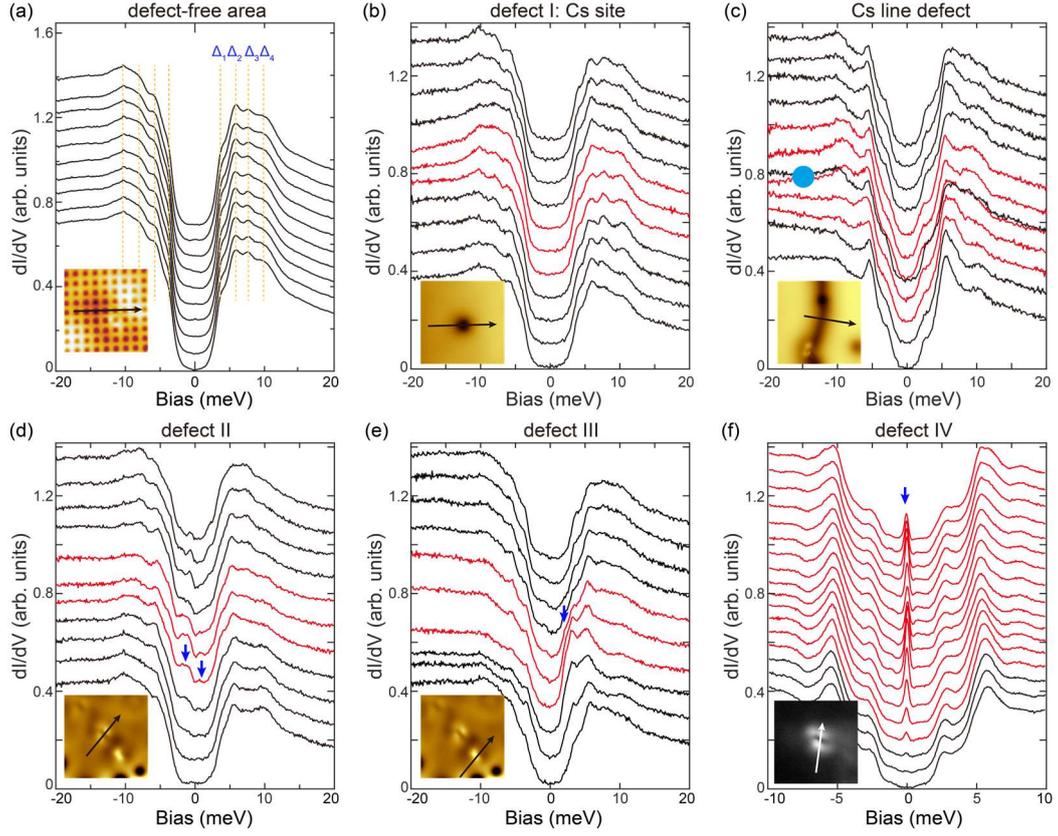

**FIG. 2. Defect effects on the superconductivity of CsCa$_2$Fe$_4$As$_4$F$_2$.** (a) Spatially resolved superconducting gap spectra collected in a defect-free sample region as indicated in the inset, revealing multiple coherence peaks as highlighted by the orange dashed lines. (b)-(f) Spatially resolved d$I$/d$V$ spectra taken along the line cuts crossing various intrinsic defects as indicated in the insets, with the red curves denoting the spectra measured right on the defect centers. The blue arrows in panels (d)-(f) highlight the existence of in-gap states. The spectra in each panel are shifted vertically for clarity. Measurement conditions: (a)-(e) $V_b$ = 20 mV, $I_t$ = 300 pA, $\Delta V$ = 0.2 mV; (f) $V_b$ = 10 mV, $I_t$ = 500 pA, $\Delta V$ = 0.1 mV.

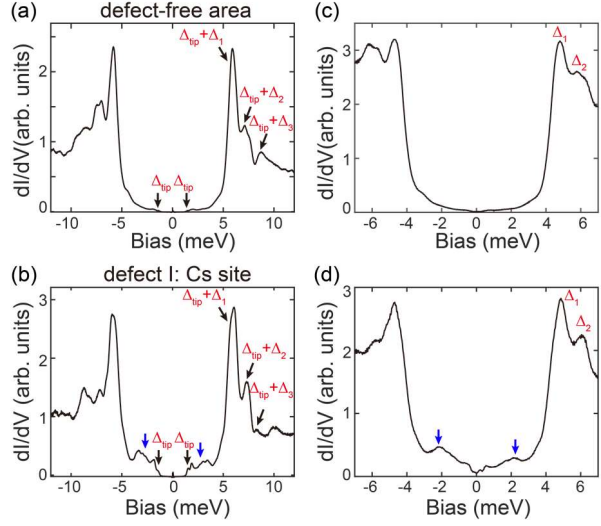

**FIG. 3. d$I$/d$V$ spectra for the defect-free area and defect I by using a superconducting Pb tip.** (a),(b) Typical d$I$/d$V$ spectra acquired with a superconducting Pb tip for the defect-free area and defect I, respectively. (c),(d) Deconvoluted spectra for panels (a) and (b). The black arrows indicate the positions of multiple coherence peaks of the superconducting Pb tip and the samples, while the blue arrows highlight the emergence of in-gap bound states for defect I. Measurement conditions: (a) $V_b$ = 12 mV, $I_t$ = 6 nA, $\Delta V$ = 0.03 mV; (c) $V_b$ = 12 mV, $I_t$ = 6 nA, $\Delta V$ = 0.02 mV.

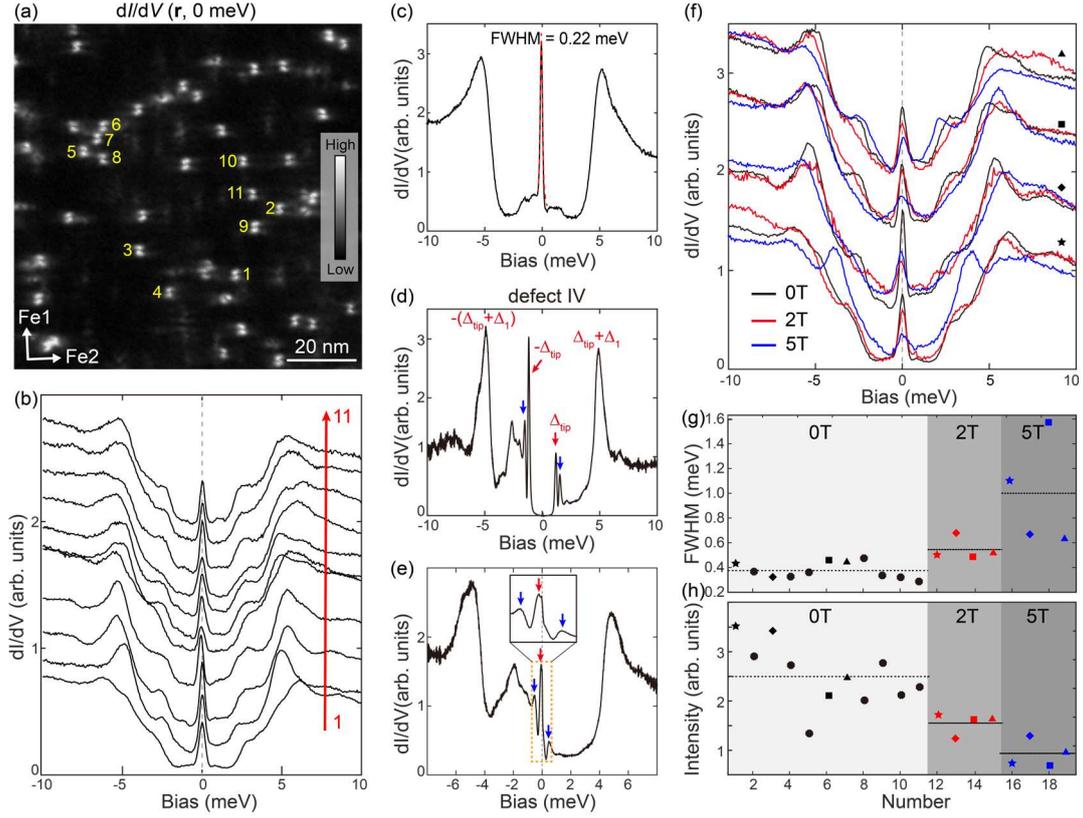

**FIG. 4. Ubiquitous ZECPs on defect IV and their magnetic field responses.** (a) Zero-energy d$I$/d$V$ map for a selected sample region, revealing the existence of numerous defects of type IV, which manifest as the bright dumbbell-like features. (b) A series of d$I$/d$V$ spectra collected from 11 individual defects of type IV as labeled in panel (a), demonstrating the consistent presence of a sharp ZECP. (c) d$I$/d$V$ spectrum for a selected defect IV, exhibiting a sharp ZECP with the FWHM of ~0.22 meV. (d) d$I$/d$V$ spectrum acquired with a superconducting Pb tip for defect IV, with the red and blue arrows indicating the positions of $\pm\Delta_{tip}$ and emerging side peaks. (e) Deconvolution of the spectrum in panel (d) to extract the intrinsic spectral feature of defect IV, where the so-called ZECP deviates from zero energy and two side peaks emerge, as highlighted in the inset. (f) Magnetic field-dependent d$I$/d$V$ spectra for four defects of type IV. (g),(h) Statistics of the FWHM and intensity of the ZECPs as a function of magnetic field. The data collected from the same defect under different magnetic fields are marked with the same symbol. Measurement conditions: (a) $V_b$ = 20 mV, $I_t$ = 350 pA, $\Delta V$ = 0.3 mV; (b)-(c) $V_b$ = 10 mV, $I_t$ = 500 pA, $\Delta V$ = 0.1 mV; (d) $V_b$ = 12 mV, $I_t$ = 6000 pA, $\Delta V$ = 0.02 mV; (f) $V_b$ = 10 mV, $I_t$ = 500 pA, $\Delta V$ = 0.1 mV.

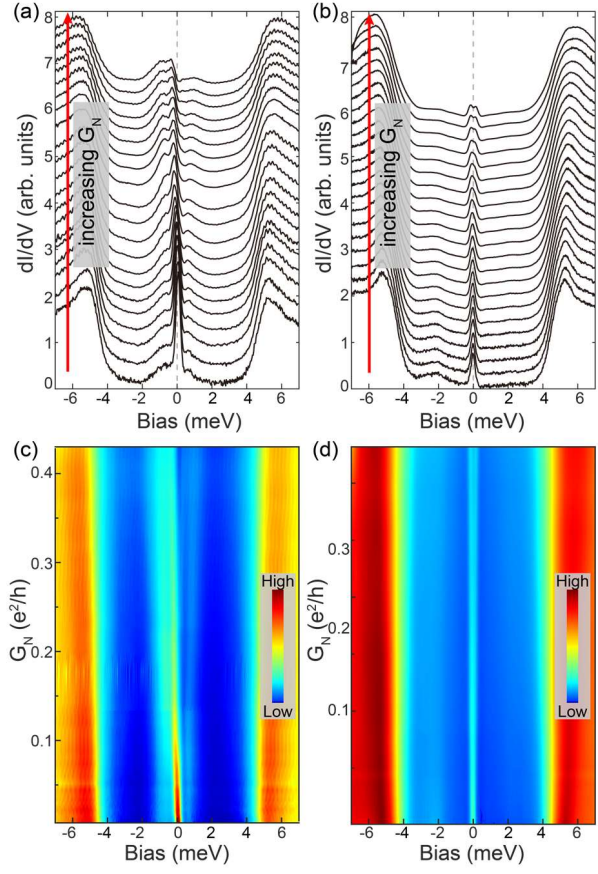

**FIG. 5. Tunneling transmissivity dependent d*I*/d*V* spectra for defect IV.** (a),(b) Tunneling transmissivity dependent d*I*/d*V* spectra for two individual defects of type IV by using a metallic Pt-Ir tip. (c),(d) Color plots of panels (a) and (b). Measurement conditions: $V_b$ = 7 mV, $I_t$ = 2-116 nA, $\Delta V$ = 0.1 mV.

# Appendix

## APPENDIX A: Characterizing the superconducting Pb tip on a Pb (111) surface

To acquire a superconducting Pb tip, we indent metallic Pt-Ir tips into a clean Pb (111) surface, which is a two-band superconductor with $\Delta_{Pb} \sim 1.3$ meV. The indentations are usually repeated many times until two conditions are satisfied: (i) the tip is single-ended; (ii) the obtained superconducting gap spectra on Pb (111) surface show sharp coherence peaks and a gap size of approximately twice of $\Delta_{Pb}$. Figure 6 presents the representative topographic image and superconducting gap spectrum obtained on Pb (111) surface with the Pb tip used in this study. The individual point defects in the topographic image are very round (Fig. 6(a)), and the spectrum exhibits a fully-developed superconducting gap with two pairs of coherent peaks at ~ 2.64 meV and 2.77 meV (red arrows in Fig. 6(b)), suggesting high quality of the superconducting Pb tip. Due to the usage of a superconducting tip, the two superconducting gaps of Pb can be resolved clearly and the coherence peaks here are much sharper than those by using a metallic tip. This unique high-resolution characteristic of the superconducting tip lays the foundation for further studies of fine spectroscopic signals.

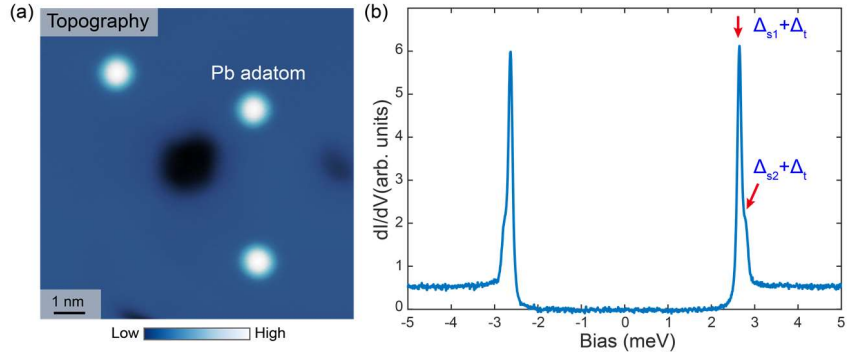

**FIG. 6. Representative topographic image and superconducting gap spectrum on Pb (111) surface acquired by the Pb tip used in this study.** Measurement conditions: (a) $V_b = 80$ mV, $I_t = 30$ pA; (b) $V_b = 5$ mV, $I_t = 100$ pA, $\Delta V = 0.02$ mV.

# APPENDIX B: Additional d$I$/d$V$ maps under more energies

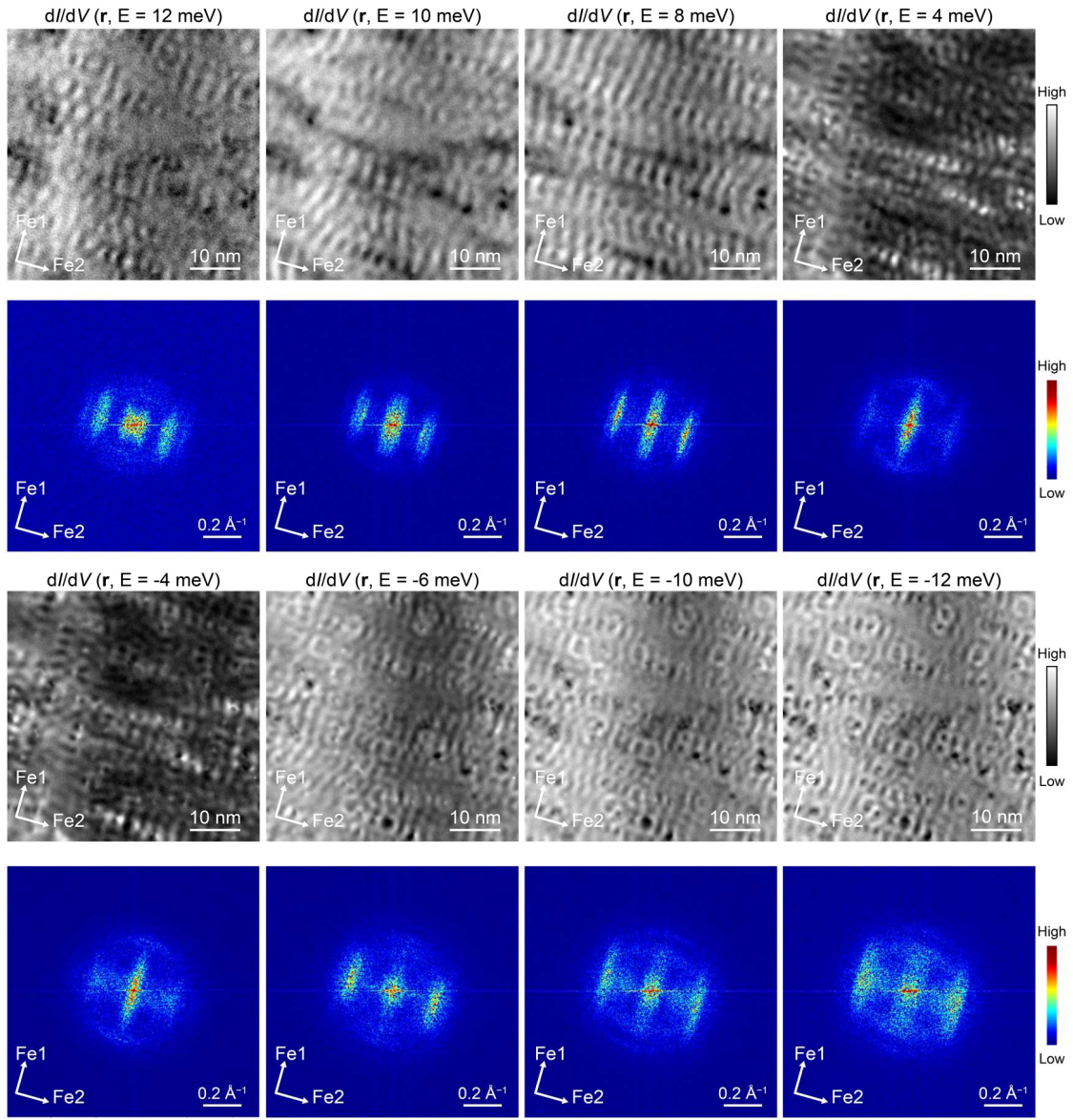

**FIG. 7. Additional d$I$/d$V$ maps and the corresponding FFT images under more energies collected in the same region of Fig. 1(c).** Measurement conditions: $V_b$ = 20 mV, $I_t$ = 300 pA, $\Delta V$ = 0.3 mV.

# APPENDIX C: Deconvolution process for the superconducting gap spectra measured using a superconducting Pb tip

In a superconductor-insulator-superconductor (SIS) tunneling junction, the measured $dI/dV$ spectrum reflects a convolution of the DOS of the tip and the sample, which is further broadened by thermal effect, quasiparticle lifetime, etc. To extract the intrinsic quasiparticle DOS of the sample, it is essential to remove the contribution of the superconducting tip from the measured spectra. Therefore, a reliable deconvolution procedure requires an accurate and independent determination of the tip's DOS.

To this end, we first calibrate the superconducting tip on a well-characterized reference system, Pb (111) single crystal, whose quasiparticle DOS is known and can be quantitatively described. In this SIS junction, both the tip and the sample are superconducting Pb, allowing the measured $dI/dV$ spectra to be directly fitted using a convolution formalism with well-defined DOS function models [29].

The differential conductance is modeled by a standard SIS tunneling expression, in which the tunneling current is determined by the convolution of the DOS of the tip and the sample weighted by the Fermi-Dirac distribution, as shown by the following formula:

$$\frac{dI}{dV}(V,T) = \frac{G_N}{e} \int_{-\infty}^{+\infty} \left\{ \frac{\partial D_t(E+eV)}{\partial V}[f(E,T)-f(E+eV,T)] - \frac{D_t(E+e)\partial f(E+eV)}{\partial V} \right\} D_s(E) dE \quad (1)$$

where $G_N$ is the normal-state conductance, $f(E,T)$ is the Fermi-Dirac distribution at temperature $T$ and energy $E$, $D_{t(s)}$ represent the DOS of the tip and the sample. Considering that bulk Pb has two superconducting gaps, we use a modified two-gap Dynes formula; while for the Pb tip, the amorphous atomic structure induces strong interband scattering, allowing its DOS to be described by a single-gap Dynes model:

$$D_s = w_{s,1} Re\left[sgn(E)\frac{E}{\sqrt{E^2+2i\gamma E-\Delta_{s,1}^2}}\right] + w_{s,2} Re\left[sgn(E)\frac{E}{\sqrt{E^2+2i\gamma E-\Delta_{s,2}^2}}\right] \quad (2)$$

$$D_t = Re\left[sgn(E)\frac{E}{\sqrt{E^2+2i\gamma E-\Delta_t^2}}\right] \quad (3)$$

where $w_{s,1(2)}$ are the weighting factors of the two gaps and satisfy $w_{s,1} + w_{s,2} = 1$. $D_{t(s)}$ are fitted with these two formulas.

By fitting the $dI/dV$ spectra measured on Pb (111) surface using a superconducting Pb tip (Fig. 8(a)), all relevant parameters, including the superconducting gap sizes, the Dynes broadening parameter $\gamma$, and the effective temperature $T_{eff}$ of the junction, are determined, which enables reliable reconstruction of DOS of the tip. The detailed fitting curve and fitting parameters are shown in Fig. 8(a), and the reconstructed superconducting gap spectrum for the Pb tip is illustrated by the orange curve in Fig. 8(b).

Once the tip's DOS is established, it is regarded as an invariant in subsequent studies. For samples with unknown electronic structures, the intrinsic DOS of the sample is obtained by numerically deconvolving the calibrated tip's DOS from the measured spectra (please see ref. [52] for more details). The validity of this deconvolution procedure is confirmed on Pb (111), where the deconvolved DOS correctly reproduces the two superconducting gaps of ~1.26 meV and 1.41 meV (blue curve in Fig. 8(b)). In this study, the same reconstructed DOS of the tip is also applied to the deconvolution procedure of the spectra acquired on CsCa$_2$Fe$_4$As$_4$F$_2$, and as discussed in the main text, the deconvolved DOS of CsCa$_2$Fe$_4$As$_4$F$_2$ is very similar to that measured with a metallic tip, except that a higher energy resolution is obtained.

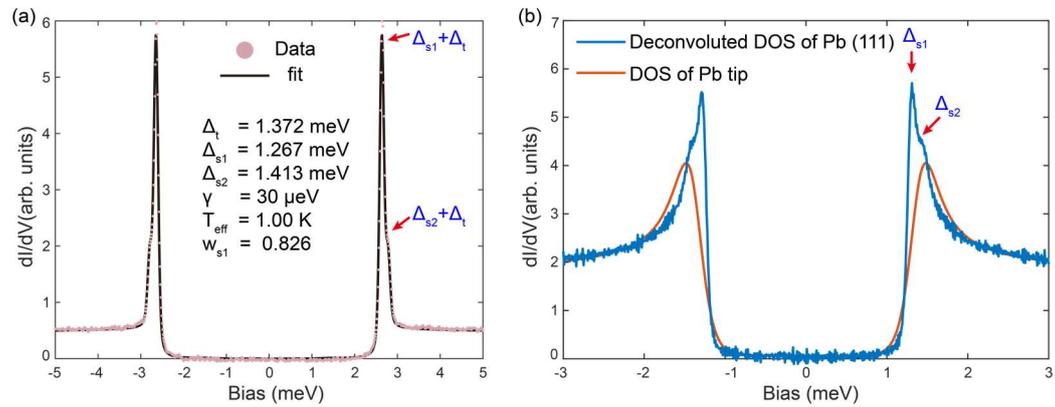

**FIG. 8. Deconvolution of the quasiparticle spectra for Pb (111).** (a) Representative superconducting gap spectrum of Pb (111) surface acquired by the Pb tip, which is the same as that in Fig. 6(b), and the black line shows the fitting results. (b) The deconvoluted DOS for Pb (111) via a numerical deconvolution procedure. Measurement conditions: $V_b$ = 5 mV, $I_t$ = 100 pA, $\Delta V$ = 0.02 mV.

**References**

[1] J. X. Yin, Z. Wu, J. H. Wang, Z. Y. Ye, J. Gong, X. Y. Hou, L. Shan, A. Li, X. J. Liang, X. X. Wu *et al.*, Nat. Phys. **11**, 543 (2015).

[2] Q. Liu, C. Chen, T. Zhang, R. Peng, Y. J. Yan, C. H. P. Wen, X. Lou, Y. L. Huang, J. P. Tian, X. L. Dong *et al.*, Phys. Rev. X **8**, 041056 (2018).

[3] P. Zhang, K. Yaji, T. Hashimoto, Y. Ota, T. Kondo, K. Okazaki, Z. Wang, J. Wen, G. D. Gu, H. Ding *et al.*, Science **360**, 182 (2018).

[4] D. Wang, L. Kong, P. Fan, H. Chen, S. Zhu, W. Liu, L. Cao, Y. Sun, S. Du, J. Schneeloch *et al.*, Science **362**, 333 (2018).

[5] S. Zhu, L. Kong, L. Cao, H. Chen, M. Papaj, S. Du, Y. Xing, W. Liu, D. Wang, C. Shen *et al.*, Science **367**, 189 (2020).

[6] C. Liu, C. Chen, X. Liu, Z. Wang, Y. Liu, S. Ye, Z. Wang, J. Hu, and J. Wang, Sci. Adv. **6**, eaax7547 (2020).

[7] P. Fan, F. Yang, G. Qian, H. Chen, Y. Y. Zhang, G. Li, Z. Huang, Y. Xing, L. Kong, W. Liu *et al.*, Nat. Commun. **12**, 1348 (2021).

[8] W. Liu, L. Cao, S. Zhu, L. Kong, G. Wang, M. Papaj, P. Zhang, Y. B, Liu, H. Chen, G. Li *et al.*, Nat. Commun. **11**, 5688 (2020).

[9] L. Cao, G. Li, W. Liu, Y. B. Liu, H. Chen, Y. Xing, L. Kong, F. Yang, Q. Hu, M. Li *et al.*, Chin. Phys. Lett. **41**, 117401 (2024).

[10] Z. Wang, J. O. Rodriguez, L. Jiao, S. Howard, M. Graham, G. D. Gu, T. L. Hughes, D. K. Morr, and V. Madhavan, Science **367**, 104 (2020).

[11] S. S. Zhang, J. X. Yin, G. Dai, L. Zhao, T. R. Chang, N. Shumiya, K. Jiang, H. Zheng, G. Bian, D. Multer *et al.*, Phys. Rev. B **101,** 100507 (2020).

[12] M. Li, G. Li, L. Cao, X. Zhou, X. Wang, C. Jin, C. K. Chiu, S. J. Pennycook, Z. Wang and H. J. Gao, Nature **606**, 890 (2022).

[13] L. Kong, L. Cao, S. Zhu, M. Papaj, G. Dai, G. Li, P. Fan, W. Liu, F. Yang, X. Wang *et al.*, Nat. Commun. **12**, 4146 (2021).

[14] X. B. Ma, G. W. Wang, R. Liu, T. Y. Yu, Y. R. Peng, P. Y. Zheng, and Z. P. Yin, Phys. Rev. B **106**, 115114 (2022).

[15] Y. Fasano, I. Maggio-Aprile, N. D. Zhigadlo, S. Katrych, J. Karpinski, and Ø. Fischer, Phys. Rev. Lett. **105**, 167005 (2010).

[16] L. Shan, Y. L. Wang, B. Shen, B. Zeng, Y. Huang, A. Li, D. Wang, H. Yang, C. Ren, Q. H. Wang *et al.*, Nat. Phys. **7**, 325 (2011).

[17] Y. Yin, M. Zech, T. L. Williams, X. F. Wang, G. Wu, X. H. Chen, and J. E. Hoffman, Phys. Rev. Lett. **102**, 097002 (2009).

[18] D. T. Adroja, F. K. K. Kirschner, F. Lang, M. Smidman, A. D. Hillier, Z. C. Wang, G. H. Cao, G. B. G. Stenning, and S. J. Blundell, J. Phys. Soc. Jpn. **87,** 124705 (2018).

[19] F. K. K. Kirschner, D. T. Adroja, Z. C. Wang, F. Lang, M. Smidman, P. J. Baker, G. H. Cao, and S. J. Blundell, Phys. Rev. B **97**, 060506(R) (2018).

[20] M. Smidman, F. K. K. Kirschner, D. T. Adroja, A. D. Hillier, F. Lang, Z. C. Wang, G. H. Cao, and S. J. Blundell, Phys. Rev. B **97**, 060509(R) (2018).

[21] T. Wang, J. Chu, J. Feng, L. Wang, X. Xu, W. Li, H. Wen, X. Liu, and G. Mu, Sci. China Phys. Mech. Astron. **63**, 297412 (2020).

[22] D. Torsello, E. Piatti, G. A. Ummarino, X. Yi, X. Xing, Z. Shi, G. Ghigo, and D. Daghero, npj

Quantum Mater. **7**, 10 (2022).

[23] E. Piatti, D. Torsello, G. Ghigo, and D. Daghero, Low Temp. Phys. **49**, 770 (2023).

[24] Y. Y. Huang, Z. C. Wang, Y. J. Yu, J. M. Ni, Q. Li, E. J. Cheng, G. H. Cao, and S. Y. Li, Phys. Rev. B **99**, 020502(R) (2019).

[25] D. S. Wu, W. Hong, C. Dong, X. Wu, Q. Sui, J. Huang, Q. Gao, C. Li, C. Song, H. Luo *et al.*, Phys. Rev. B **101**, 224508 (2020).

[26] P. Li, S. Liao, Z. Wang, H. Li, S. Su, J. Zhang, Z. Chen, Z. Jiang, Z. Liu, L. Yang *et al.*, Nat. Commun. **15**, 6433 (2024).

[27] W. Duan, K. Chen, W. Hong, X. Chen, H. Yang, S. Li, H. Luo, and H. H. Wen, Phys. Rev. B **103**, 214518 (2021).

[28] Z. C. Wang, C. Y. He, S. Q. Wu, Z. T. Tang, Y. Liu, A. Ablimit, C. M. Feng, and G. H. Cao, J. Am. Chem. Soc. **138**, 7856 (2016).

[29] D. Cho, K. M. Bastiaans, D. Chatzopoulos, G. D. Gu, and M. P. Allan, Nature **571**, 541 (2019).

[30] S. Shao, F. Zhang, Z. Zhang, T. Wang, Y. Wu, Y. Tu, J. Hou, X. Hou, N. Hao, G. Mu *et al.*, Sci. China Phys. Mech. Astron. **66**, 287412 (2023).

[31] W. Zeng, Z. Zhang, X. Dong, Y. Tu, Y. Wu, T. Wang, F. Zhang, S. Shao, J. Hou, X. Hou *et al.*, Chin. Phys. B **34**, 087402 (2025).

[32] C. L. Song, Y. L. Wang, P. Cheng, Y. P. Jiang, W. Li, T. Zhang, Z. Li, K. He, L. Wang, J. F. Jia *et al.*, Science **332**, 1410 (2011).

[33] T. Zhang, Y. Hu, W. Su, C. Chen, X. Wang, D. Li, Z. Lu, W. Yang, Q. Zhang, X. Dong *et al.*, Phys. Rev. Lett. **130**, 206001 (2023).

[34] W. Li, H. Ding, P. Deng, K. Chang, C. Song, K. He, L. Wang, X. Ma, J. P. Hu, X. Chen *et al.*, Nat. Phys. **8**, 126 (2012).

[35] S. Grothe, S. Chi, P. Dosanjh, R. Liang, W. N. Hardy, S. A. Burke, D. A. Bonn, and Y. Pennec, Phys. Rev. B **86**, 174503 (2012).

[36] V. Grinenko, K. Kikoin, S.-L. Drechsler, G. Fuchs, K. Nenkov, S. Wurmehl, F. Hammerath, G. Lang, H.-J. Grafe, B. Holzapfel, et al., Phys. Rev. B **84**, 134516 (2011).

[37] W. Chen, Z. Tian, P. Li, W. Luo, and C. L. Gao, Phys. Rev. B **96**, 214426 (2017).

[38] S. H. Pan, E. W. Hudson, and J. C. Davis, Appl. Phys. Lett. **73**, 2992 (1998).

[39] M. Ruby, B. W. Heinrich, J. I. Pascual, and K. J. Franke, Phys. Rev. Lett. **114**, 157001 (2015).

[40] M. Ruby, F. Pientka, Y. Peng, F. von Oppen, B. W. Heinrich, and K. J. Franke, Phys. Rev. Lett. **115**, 197204 (2015).

[41] I. I. Mazin, D. J. Singh, M. D. Johannes, and M. H. Du, Phys. Rev. Lett. **101**, 057003 (2008).

[42] P. J. Hirschfeld, M. M. Korshunov, and I. I. Mazin, Rep. Prog. Phys. **74**, 124508 (2011).

[43] T. Hanaguri, S. Niitaka, K. Kuroki, and H. Takagi, Science **328**, 474 (2010).

[44] P. Dai, Rev. Mod. Phys. **87**, 855 (2015).

[45] V. Mourik, K. Zuo, S. M. Frolov, S. R. Plissard, E. P. A. M. Bakkers, and L. P. Kouwenhoven, Science **336**, 1003 (2012).

[46] K. T. Law, P. A. Lee, and T. K. Ng, Phys. Rev. Lett. **103**, 237001 (2009).

[47] B. E. Feldman, M. T. Randeria, J. Li, S. Jeon, Y. Xie, Z. Wang, I. K. Drozdov, B. Andrei Bernevig, and A. Yazdani, Nat. Phys. **13**, 286 (2017).

[48] D. Chatzopoulos, D. Cho, K. M. Bastiaans, G. O. Steffensen, D. Bouwmeester, A. Akbari, G. Gu, J. Paaske, B. M. Andersen, and M. P. Allan, Nat. Commun. **12**, 298 (2021).

[49] C. Chen, Q. Liu, T. Z. Zhang, D. Li, P. P. Shen, X. L. Dong, Z. X. Zhao, T. Zhang, and D. L.

Feng, Chin. Phys. Lett. **36**, 057403 (2019).

[50] H. Huang, R. Drost, J. Senkpiel, C. Padurariu, B. Kubala, A. L. Yeyati, J. C. Cuevas, J. Ankerhold, K. Kern, and C. R. Ast, Commun. Phys. **3**, 199 (2020).

[51] S. Karan, H. Huang, A. Ivanovic, C. Padurariu, B. Kubala, K. Kern, J. Ankerhold, and C. R. Ast, Nat. Commun. **15**, 459 (2024).

[52] A. Palacio-Morales, E. Mascot, S. Cocklin, H. Kim, S. Rachel, D. K. Morr, and R. Wiesendanger, Sci. Adv. **5**, eaav6600 (2019).